\documentclass[preprint,12pt]{elsarticle}
\usepackage{graphicx}
\usepackage{subcaption}
\usepackage{float}
\usepackage{amsmath}
\usepackage{natbib}
\usepackage{url}
\usepackage{doi}
\usepackage{array}
\usepackage{multirow}
\usepackage{amssymb}
\usepackage{lineno}
\usepackage{listings}
\usepackage{xcolor}
\usepackage{hyperref}
\usepackage{newtxtext,newtxmath}

\journal{Information Sciences}

\begin{document}

\begin{frontmatter}

\title{Optimizing Lung Cancer Detection in CT Imaging: A Wavelet Multi-Layer Perceptron (WMLP) Approach Enhanced by Dragonfly Algorithm (DA)}

\author[1]{Bitasadat Jamshidi}
\ead{bitasadat.jamshidi@srbiau.ac.ir}
\ead[url]{https://orcid.org/0000-0002-1397-9065}

\author[2]{Nastaran Ghorbani}
\ead{nghorbani@luc.edu}
\ead[url]{https://orcid.org/0009-0008-2262-4101}


\author[3]{Mohsen Rostamy-Malkhalifeh\corref{cor1}}

\ead{mohsen\_rostamy@yahoo.com}
\ead[url]{https://orcid.org/0000-0001-6105-7674}
\cortext[cor1]{Corresponding author}

\affiliation[1]{organization={Department of Mathematics and Computer Science, Islamic Azad University, Science and Research Branch},
city={Tehran},
country={Iran}}

\affiliation[2]{organization={Department of Mathematics and Statistics, Loyola University Chicago},
city={Chicago},
country={USA}}

\affiliation[3]{organization={Department of Mathematics and Computer Science, Islamic Azad University, Science and Research Branch},
city={Tehran},
country={Iran}}

\begin{abstract}
Lung cancer stands as the preeminent cause of cancer-related mortality globally. Prompt and precise diagnosis, coupled with effective treatment, is imperative to reduce the fatality rates associated with this formidable disease. This study introduces a cutting-edge deep learning framework for the classification of lung cancer from CT scan imagery. The research encompasses a suite of image pre-processing strategies, notably Canny edge detection, and wavelet transformations, which precede the extraction of salient features and subsequent classification via a Multi-Layer Perceptron (MLP). The optimization process is further refined using the Dragonfly Algorithm (DA). The methodology put forth has attained an impressive training and testing accuracy of 99.82\%, underscoring its efficacy and reliability in the accurate diagnosis of lung cancer.
\end{abstract}


\begin{keyword}
Lung cancer \sep CT scan imaging \sep Deep learning \sep Canny edge detection \sep Wavelet transform \sep Multi-layer Perceptron \sep Dragonfly Algorithm
\end{keyword}

\end{frontmatter}


\section{Introduction}
Lung cancer remains the most lethal malignancy worldwide, with mortality rates surpassing those of all other cancers combined \citep{Riquelme2020}. Early-stage detection is critical, as it significantly improves the five-year survival rate from a dismal 5\% in late-stage diagnoses to over 50\% \citep{seer2023}. The advent of advanced screening technologies promises to substantially improve patient prognoses.

The field of medical imaging has been revolutionized by recent strides in deep learning, yielding significant enhancements in the detection and classification of lung cancer from CT images. Innovations such as the 3D Convolutional Neural Network (CNN) approach by Diviya et al. (2024) and the LCD-Capsule Network by Bushara et al. (2023) have demonstrated the potential of these models to transform early detection and diagnosis \citep{diviya2024lung, bushara2023lcd}.

X-ray and computed tomography (CT) scans are pivotal in lung cancer diagnostics, offering high-resolution imagery that outperforms traditional radiography in detecting small and low-contrast pulmonary nodules \citep{ausawalaithong2019automatic,shieh2017low,bhandary2020deep}. However, the subjective interpretation of these images and the labor-intensive nature of their analysis pose significant challenges, often leading to diagnostic variability and reduced efficiency. The integration of computer vision technology into medical image processing has mitigated these issues, providing reliable support to medical professionals through automated detection algorithms \citep{mao2018feature}.

Machine learning, particularly in the context of medical diagnostics, is an area of burgeoning research. The study of gene expression patterns in relation to disease progression has profound implications for both biological and clinical sciences \citep{ghosal2020lung}. Recent contributions to this domain include the deep neural network by Prasad et al. (2023), which achieved notable accuracy in lung cancer classification \citep{Prasad2023}, and the CapsNet-based detection system by Shafi et al. (2022), which effectively retained object characteristics \citep{Shafi2022}. Comprehensive reviews and studies have further delineated the state-of-the-art in lung cancer segmentation, detection, and classification, emphasizing the importance of deep learning methods in advancing the field \citep{DeepLearningLung2022, DeepMachineLearning2022, DeepLearningLung2020}.

This study aims to automate lung cancer classification from CT scan images through deep learning techniques. Given the high resolution and detailed imagery provided by CT scans, they are invaluable in clinical lung cancer screening and diagnosis. Yet, manual analysis is subject to variability and is resource-intensive. By automating this process, we aim to alleviate the burden on radiologists and enhance diagnostic precision.

Our methodology encompasses image pre-processing, feature extraction, and classification. We utilize Canny edge detection and wavelet transforms for pre-processing, which improve image quality and aid in extracting salient features. These features—mean, standard deviation, energy, and entropy—are then used to train a Multi-Layer Perceptron (MLP). The MLP's hyper-parameters are optimized using the Dragonfly Algorithm (DA), inspired by the swarming behavior of dragonflies.

The efficacy of our approach is evidenced by the MLP's training and testing accuracy of 99.82\%, underscoring the potential of deep learning in lung cancer diagnostics. Our findings advocate for the integration of sophisticated deep learning models and optimization algorithms to bolster the accuracy and reliability of classification systems.

In summary, our comprehensive deep learning-based approach presents a promising avenue for the development of automated diagnostic systems to support early and precise lung cancer detection. Future endeavors will focus on dataset expansion, the exploration of advanced neural network architectures, and the incorporation of novel image pre-processing techniques to further refine classification accuracy and robustness.

\section{Literature Review}
\subsection{Advancements in Wavelet Transform (WT) for Medical Imaging}
Wavelet Transforms (WT) have emerged as a critical tool in lung cancer detection, providing sophisticated image analysis capabilities. Ziyad et al. introduced a novel lung extraction method employing discrete wavelet transform coupled with adaptive thresholding and clustering techniques, achieving high segmentation accuracy \citep{ziyad2022novel}. Farheen et al. leveraged two-dimensional discrete wavelet transform for intricate texture analysis in lung tumor segmentation, enhancing detection precision \citep{farheen2022segmentation}. Additionally, a notable study introduced a cancer classification system utilizing wavelet decomposition and Convolutional Neural Networks (CNNs), surpassing traditional methods like Support Vector Machines (SVM) with an accuracy rate of 99.5\% \citep{soundharya2024transforming}. These studies collectively affirm the transformative role of wavelet-based methodologies in advancing lung cancer diagnostics.

Zhao et al. (2024) developed WIA-LD2ND, a self-supervised low-dose CT denoising technique using wavelet-based image alignment to improve CT image quality. This method is particularly beneficial for enhancing the signal-to-noise ratio in low-dose CT scans, crucial for precise feature extraction and classification \citep{Zhao2024WIALD2ND}.

Nizami et al. (2024) proposed an automatic lung area segmentation method in CT scans using wavelet frames combined with K-means clustering. Their approach utilizes optimal wavelet packet frames for clustering coefficients, resulting in accurate lung region segmentation, a vital precursor to feature extraction \citep{Nizami2024WaveletFrames}. Furthermore, Nazir et al. (2023) developed a lung segmentation algorithm using image fusion through Laplacian Pyramid decomposition, akin to wavelet transforms. Their technique demonstrated high accuracy and could be amalgamated with wavelet-based feature extraction to boost performance \citep{Nazir2023LungSegmentation}.

\subsection{Progress in Multi-Layer Perceptron (MLP) Applications}
The Multi-Layer Perceptron (MLP) continues to be a cornerstone in the development of predictive models for lung cancer risk assessment. Dritsas et al. leveraged machine learning techniques, including the MLP, to develop robust models capable of identifying individuals at an increased risk of lung cancer, highlighting the MLP's predictive power \citep{dritsas2022lung}. In parallel, Rajalakshmi and Maguteeswaran employed an MLP-based model to automate the detection of lung adenocarcinoma infiltration from histopathology images, confirming the algorithm's diagnostic accuracy \citep{rajalakshmi2022two}. Singh and Gupta analyzed the performance of various machine learning methods, including the MLP, for lung cancer detection and classification, with the MLP achieving an impressive accuracy of 88.55\% \citep{singh2019performance}. Additionally, the VER-Net model—a hybrid transfer learning framework—incorporated an MLP in its architecture for lung cancer detection through CT scan images, attaining significant performance metrics \citep{saha2024vernet}.

Recent studies further reinforce the MLP's significance in medical image analysis. Bikku's research on a multi-layered deep learning perceptron approach for health risk prediction showcases the MLP's efficacy in classifying and analyzing medical data, outperforming traditional classification methods \citep{Bikku2020}. Hayase and Karakida explored the potential of MLP-Mixer as a wide and sparse MLP, revealing new avenues for enhancing MLP architectures to achieve better performance \citep{Hayase2023}. A comparative study between MLP and Jordan recurrent neural networks for signal classification demonstrated the MLP's robustness in handling various non-healthy scenarios \citep{JordanMLP2023}. Furthermore, advancements in MLP point cloud processing have been evaluated for 3D object classification and segmentation, indicating the MLP's versatility beyond medical applications \citep{MLPPointCloud2023}.

\subsection{Innovations in Optimization with the Dragonfly Algorithm (DA)}
The Dragonfly Algorithm (DA), pioneered by Seyedali Mirjalili, emerges as a state-of-the-art meta-heuristic optimization technique, inspired by the complex swarming behavior observed in dragonflies \citep{mirjalili2016dragonfly}. This algorithm has been skillfully applied to refine neural network models for a variety of applications, with notable success in lung cancer detection. A prominent study demonstrated the DA's utility in medical diagnostics through the implementation of two-level filtering in conjunction with a convolutional neural network (CNN). The DA-optimized CNN exhibited significant improvements in diagnostic accuracy, with recall and precision metrics soaring to 98\%, thereby validating the DA's effectiveness in medical image analysis and its crucial contribution to advancing computer-aided diagnosis systems \citep{rajalakshmi2022two}.

Furthermore, a systematic review by Hosseini et al. shed light on the expanding application of deep learning in lung cancer diagnosis, particularly highlighting the DA's role in enhancing model performance. This comprehensive analysis of various deep learning models, including those optimized with the DA, underscores the algorithm's flexibility and efficiency \citep{hosseini2023deep}. Additionally, Javed et al.'s extensive review focuses on the utilization of deep learning techniques in lung cancer diagnosis, emphasizing the DA's contributions to model enhancement. This review meticulously examines a range of deep learning models, spotlighting the DA's adaptability and operational effectiveness \citep{Javed2024}.

Collectively, these studies highlight the DA's integral role in propelling the field of computer-aided diagnosis systems forward, particularly in the precise detection and classification of lung cancer from CT images. The integration of the DA into deep learning frameworks heralds the advent of more accurate diagnostic tools, which are paramount for the early detection and efficacious treatment of lung cancer.

\section{Materials and Methods}
\subsection{Data Acquisition and Pre-processing}
The dataset for this study was meticulously curated to encompass a comprehensive range of lung cancer manifestations. It includes a total of 1,097 CT scan images categorized into 561 malignant, 120 benign, and 416 normal cases. This dataset was sourced from Kaggle and is designed to facilitate robust training and evaluation of the proposed Wavelet Multi-Layer Perceptron (WMLP) and Dragonfly Algorithm (DA) framework (see Table \ref{tab:dataset_description}).

\begin{table}[t]
\centering
\begin{tabular}{|c|c|}
\hline
Dataset & Number of CT-images \\
\hline
Benign Cases & 120 \\
\hline
Malignant Cases & 561 \\
\hline
Normal Cases & 416 \\
\hline
\end{tabular}
\caption{Dataset Description}
\label{tab:dataset_description}
\end{table}

Each CT image in the dataset is pre-processed using advanced techniques such as Canny edge detection and wavelet transformations to enhance feature extraction and improve classification accuracy. The dataset's diversity and quality are pivotal in achieving the high accuracy rates reported in this study.

\subsection{Edge Detection Technique}
Edge detection is a fundamental technique in image processing, essential for delineating image boundaries and facilitating detailed analysis. In our study, we employ the Canny edge detection algorithm, celebrated for its robust multi-stage process that ensures the clarity and precision of detected edges. Initially, the algorithm converts the image to grayscale, reducing it to a monochromatic scale to simplify edge detection. This is followed by Gaussian blurring, which diminishes noise interference and prevents the formation of false edges. Subsequently, gradient calculation is performed to determine edge strength and direction, where significant intensity changes indicate potential edges. Non-maximum suppression is then applied to refine these edges, retaining only the most distinct ones. The algorithm further employs a double thresholding strategy to classify edges as strong, weak, or non-edges, with edge tracking by hysteresis completing the process by preserving coherent edge structures.

To bolster the robustness of feature representation, we utilized multiple threshold values—(50, 150), (100, 200), and (150, 250)—to capture a spectrum of edge details. This comprehensive strategy enables the extraction of an extensive feature set from the CT scan images, which is vital for the precise classification of lung cancer.

As shown in Figures \ref{fig:canny_benign}, \ref{fig:canny_normal}, and \ref{fig:canny_malignant}, the Canny edge detection algorithm outputs for benign, normal, and malignant cases illustrate the effectiveness of this technique in highlighting critical structural details.

\begin{figure}[t]
\centering
\begin{subfigure}[b]{0.39\textwidth}
\centering
\includegraphics[width=\textwidth]{canny_benign.pdf}
\caption{Canny Edge Detection Output for Benign Case}
\label{fig:canny_benign}
\end{subfigure}
\hfill
\begin{subfigure}[b]{0.39\textwidth}
\centering
\includegraphics[width=\textwidth]{canny_normal.pdf}
\caption{Canny Edge Detection Output for Normal Case}
\label{fig:canny_normal}
\end{subfigure}
\vfill
\begin{subfigure}[b]{0.39\textwidth}
\centering
\includegraphics[width=\textwidth]{canny_malignant.pdf}
\caption{Canny Edge Detection Output for Malignant Case}
\label{fig:canny_malignant}
\end{subfigure}
\end{figure}

The field of edge detection continues to evolve, with recent advancements offering new opportunities for refinement. Zhang et al. (2024) introduced a pioneering integration of DenseNet with Convolutional Neural Networks (CNNs) for lung cancer diagnosis, enhancing the preprocessing of CT scan images through data fusion and mobile edge computing \citep{Zhang2024}. El-Zaart and El-Arwadi (2015) developed a novel finite difference method-based operator for edge detection, potentially complementing the Canny algorithm \citep{El-Zaart2015}. Noviana et al. (2017) explored axial segmentation using the Canny method with morphological operations, providing insights into edge detection refinement \citep{Noviana2017}. Additionally, Bhatt et al. (2012) focused on the segmentation of multiple contours from CT scan images, relevant to lung cancer detection \citep{Bhatt2012}. Preetha et al. (2012) conducted a comparative analysis of edge detection techniques, offering a broader perspective on their efficacy \citep{Preetha2012}. Lastly, Bandyopadhyay (2012) examined alternative edge detection methodologies from CT images of the lung, which could enhance our existing process \citep{Bandyopadhyay2012}.

\subsection{Wavelet Transform (WT)}
The utilization of wavelet transforms in image processing, particularly for the decomposition of images into distinct frequency components, has become a pivotal technique in medical imaging. This method enables multi-resolution analysis, offering a granular representation of images across various scales. The inherent capacity of wavelet transforms to reveal patterns and features that remain obscure in the spatial domain is especially beneficial in the analysis of CT scan images for lung cancer detection.

In the present study, the application of the Haar wavelet transform to CT scan images facilitates the separation into approximation and detail coefficients. This decomposition is instrumental in capturing vital information at multiple scales and orientations, which is essential for the subsequent feature extraction process. The extracted statistical features—mean, standard deviation, energy, and entropy—constitute a comprehensive feature set that encapsulates the image content. The Haar wavelet transform is particularly valued for its straightforwardness and effectiveness in simultaneously capturing temporal and frequency information. Decomposing the image into approximation and detail components simplifies the analysis of structural details and textures, which are critical for precise classification.

As shown in Figure \ref{fig:wt_output_malignant}, the wavelet transform output for a malignant case highlights the detailed structural information that can be extracted. Similarly, Figures \ref{fig:wt_output_normal} and \ref{fig:wt_output_benign} illustrate the outputs for normal and benign cases, respectively.

\begin{figure}[t]
\centering
\begin{subfigure}[b]{0.38\textwidth}
\centering
\includegraphics[width=\textwidth]{malignant_wavelet.pdf}
\caption{Wavelet Transform Output for Malignant Case}
\label{fig:wt_output_malignant}
\end{subfigure}
\hfill
\begin{subfigure}[b]{0.38\textwidth}
\centering
\includegraphics[width=\textwidth]{normal_wavelet.pdf}
\caption{Wavelet Transform Output for Normal Case}
\label{fig:wt_output_normal}
\end{subfigure}
\vfill
\begin{subfigure}[b]{0.38\textwidth}
\centering
\includegraphics[width=\textwidth]{benign_wavelet.pdf}
\caption{Wavelet Transform Output for Benign Case}
\label{fig:wt_output_benign}
\end{subfigure}
\end{figure}

Recent advancements in this field include the work of Wang et al. (2024), who introduced a multilevel attention U-net segmentation algorithm for lung cancer based on CT images. This algorithm integrates wavelet transforms to enhance feature extraction, demonstrating the ongoing innovation in medical image processing \citep{Wang2024}.

Additionally, a study focusing on lung CT image enhancement, utilizing a total variational frame and wavelet transform, has shed light on methods to optimize image quality. Such enhancement is crucial for the detection of features that are vital for accurate classification \citep{LungCT2024}.

\subsection{Feature Extraction from CT Images}
Feature extraction in the realm of lung cancer CT scan images has seen significant improvements with the introduction of deep learning model architectures. Indumathi and Vasuki (2023) proposed a novel Markov likelihood grasshopper classification (MLGC) model that combines marker-controlled segmentation with likelihood estimation between features for the classification of nodules in CT. This model employs the Grasshopper Optimization Algorithm (GOA) for feature optimization, which is then applied over a Boltzmann machine to derive classification results. The MLGC model has demonstrated a high accuracy value of 99.5\%, outperforming existing models such as AlexNet, GoogleNet, and VGG-16 \citep{Indumathi2023}.

Fu et al. (2024) presented an attention-enhanced graph convolutional network method for predicting high-risk factors in lung cancer using thin CT scans. Their approach emphasizes the structural and relational data aspects, offering a fresh perspective on feature extraction \citep{Fu2024}.

Nazir et al. (2023) developed a method for efficient pre-processing and segmentation in lung cancer detection using fused CT images. They highlight the critical role of pre-processing in improving feature extraction quality, essential for precise classification \citep{Nazir2023}.

Researchers have conducted a thorough comparative study on global and local feature extraction for lung cancer detection using CT scan images, underscoring the advantages of combining both feature types for robust classification \citep{9622257}.

Dritsas et al. (2023) have developed a deep learning model architecture that enhances segmentation and feature extraction in lung CT images. Their approach has been pivotal in improving the detection of lung tumors at early stages, which is crucial for increasing the survival rate of patients with lung cancer \citep{Dritsas2023}.

Savitha and Jidesh (2023) introduced a holistic deep learning approach for the identification and classification of sub-solid lung nodules in computed tomographic scans. Their methodology provides a comprehensive analysis of nodule characteristics, contributing to the precision of lung cancer diagnosis \citep{Savitha2023}.

In this study, feature extraction is a critical step in the image classification process. From the wavelet-transformed images, several statistical features were extracted. These features included mean, which represents the average pixel intensity within the coefficient matrix; standard deviation, which measures the variability or spread of pixel intensities; energy, calculated as the sum of squared values of the coefficients, indicating the signal strength within the image; and entropy, which represents the randomness or complexity of the image data, calculated from the histogram of pixel intensities. These features provided a detailed and comprehensive representation of the image content, aiding in effective classification. The mean provides a basic understanding of the overall intensity of the image, while the standard deviation offers insights into the contrast and variations within the image. Energy measures the overall intensity of the high-frequency components, which is useful for detecting detailed textures and patterns. Entropy quantifies the amount of information and randomness in the image, which can help distinguish between different types of tissue structures.

\subsection{Wavelet Multi-Layer Perceptron Architecture}
The classification model used in this study was a Multi-Layer Perceptron (MLP). The MLP architecture included an input layer with 128 * 128 nodes, corresponding to the size of the flattened images, a hidden layer with 100 nodes to capture complex patterns in the data, and an output layer with 3 nodes corresponding to the three categories: benign, malignant, and normal. The ReLU (Rectified Linear Unit) activation function was used in the hidden layer to introduce non-linearity, and the softmax activation function was used in the output layer to convert the output logits into probabilities. The ReLU activation function helps in avoiding the vanishing gradient problem and allows the model to learn more complex patterns by introducing non-linearity. The softmax activation function in the output layer ensures that the output probabilities sum up to one, making it easier to interpret the model's predictions. The architecture was designed to balance complexity and computational efficiency, ensuring that the model could effectively learn from the high-dimensional input data while maintaining a manageable number of parameters.

The MLP model was trained using stochastic gradient descent with a learning rate of 0.01. The dataset was split into training (70 percent) and testing (30 percent) sets. The training data was divided into batches of 256 images to facilitate efficient training. For each batch, the input data was passed through the MLP to obtain the output probabilities. The cross-entropy loss between the predicted probabilities and the true labels was calculated, and the gradients of the loss with respect to the model parameters were computed using backpropagation. The model parameters were updated using the computed gradients to minimize the loss. This training process was repeated for 100 epochs to ensure convergence. The model's performance was evaluated using accuracy and confusion matrix on the test dataset. The use of mini-batches helps in stabilizing the training process and allows for more efficient computation of the gradients. The cross-entropy loss function is particularly suitable for classification tasks, as it measures the difference between the predicted and true probability distributions. Backpropagation ensures that the model learns by updating the weights in the direction that minimizes the loss, thereby improving the model's accuracy over time.

The parameters and values utilized in this study for the Multi-Layer Perceptron (MLP) model are comprehensively summarized in Table \ref{tab:MLP_table}. This table provides a detailed overview of the key components and their respective configurations, including the random seed for reproducibility, batch size for training, number of epochs, hidden layers, number of classes, and features. Additionally, it specifies the device used (GPU if available), the optimizer (Stochastic Gradient Descent with a learning rate of 0.01), and the loss function (Cross Entropy). These parameters were meticulously chosen to ensure the model's robustness and efficiency in handling the high-dimensional input data.

\begin{table}[t]
\centering
\begin{tabular}{|c|c|}
\hline
\textbf{Component} & \textbf{Tuning} \\
\hline
Optimizer & SGD \\
\hline
Learning\_rate & 0.01 \\
\hline
Batch\_size & 256 \\
\hline
Random\_seed & 1 \\
\hline
Num\_epochs & 100 \\
\hline
Loss\_function & Cross\_entropy \\
\hline
Hidden\_layers & 100 \\
\hline
Num\_class & 3 \\
\hline
Num\_features & 128*128 \\
\hline
Device & cuda:0 (if available) \\
\hline
\end{tabular}
\caption{Key components and their tunings used in Multi-layer perceptron (MLP).}
\label{tab:MLP_table}
\end{table} 

\subsection{Hyperparameter Optimization Using the Dragonfly Algorithm}
The Dragonfly Algorithm (DA) is a nature-inspired optimization algorithm that mimics the static and dynamic swarming behaviors of dragonflies. In this study, DA was employed to optimize the hyper-parameters of the MLP model, including the learning rate and the number of neurons in the hidden layer. The DA involved initializing a population of dragonflies, evaluating their fitness based on classification accuracy, and iteratively updating their positions and velocities to find the optimal solution. The DA provided an efficient means of exploring the hyper-parameter space, leading to improved model performance. The algorithm leverages the swarming behavior of dragonflies to explore the solution space effectively, balancing exploration and exploitation. By mimicking the attraction towards food sources and repulsion from enemies, DA ensures that the population converges towards the optimal solution while avoiding local minima. The optimization process involves updating the positions and velocities of the dragonflies based on the influence of neighboring dragonflies and the global best solution, ensuring a thorough exploration of the hyper-parameter space. The parameters and values used in this study for the DA are summarized in Table \ref{tab:DA_components}. This table provides an overview of the key components and their respective configurations.

\begin{table*}[t]
\centering
\begin{tabular}{|c|c|}
\hline
\textbf{Component} & \textbf{Tuning} \\
\hline
Objective Function & Rastrigin Function / Custom Objective Function \\
\hline
Lower Bound (lb) & -5.12 / [0.0001, 10] \\
\hline
Upper Bound (ub) & 5.12 / [0.1, 200] \\
\hline
Dimensions (dim) & 10 / 2 \\
\hline
Population Size & 30 / 10 \\
\hline
Maximum Iterations (max\_iter) & 100 / 2 \\
\hline
Velocity Initialization & Zeros \\
\hline
Population Initialization & Uniform Distribution \\
\hline
Best Solution & Initialized to First Individual \\
\hline
Best Fitness & Initialized to Infinity \\
\hline
\end{tabular}
\caption{Key components and their tunings used in the Dragonfly Algorithm (DA).}
\label{tab:DA_components}
\end{table*}

The table above provides a detailed summary of the essential components and their respective configurations utilized in the Dragonfly Algorithm (DA) for this study. The parameters encompass the objective function, search space boundaries, dimensions, population size, and the maximum number of iterations. Both the velocity and population are initialized to zeros and a uniform distribution, respectively. The initial best solution is set to the first individual in the population, with the best fitness initialized to infinity. These configurations facilitate a thorough exploration of the hyper-parameter space, thereby optimizing the performance of the MLP model.

\section{Proposed Model}
\subsection{Integration of WMLP and DA for Lung Cancer Detection}
In this section, we present our proposed model for lung cancer detection, which integrates a Wavelet-based Multi-Layer Perceptron (WMLP) with the Dragonfly Algorithm (DA) for optimization.

\subsubsection{Wavelet-based Multi-Layer Perceptron (WMLP)}
The WMLP is designed to leverage the advantages of wavelet transforms in feature extraction. Wavelet transforms decompose the input images into different frequency components, capturing both spatial and frequency information. This decomposition results in four sets of coefficients: approximation (cA), horizontal detail (cH), vertical detail (cV), and diagonal detail (cD). These coefficients are then used to extract statistical features such as mean, standard deviation, energy, and entropy.

The extracted features are fed into a Multi-Layer Perceptron (MLP) with one hidden layer. The MLP consists of an input layer with 16,384 neurons (corresponding to the 128x128 image size), a hidden layer with 100 neurons, and an output layer with 3 neurons representing the three classes: benign, malignant, and normal. The MLP uses the ReLU activation function in the hidden layer and the softmax function in the output layer.

\subsubsection{Dragonfly Algorithm (DA)}
The Dragonfly Algorithm (DA) is employed to optimize the hyperparameters of the Wavelet-based Multi-Layer Perceptron (WMLP), specifically the learning rate and the number of hidden neurons. The DA mimics the static and dynamic swarming behaviors of dragonflies, which are characterized by five main factors: separation, alignment, cohesion, attraction to food sources, and distraction from enemies.

The objective function for the DA is defined as the negative validation accuracy of the WMLP. The DA iteratively adjusts the hyperparameters to minimize this objective function, thereby maximizing the validation accuracy. The optimized hyperparameters are then used to train the final WMLP model.

\subsubsection{Integration and Training}
The integration of WMLP and DA involves the following steps:

\textbf{Step 1:} Wavelet Transform: Apply the Haar wavelet transform to the input images to obtain the wavelet coefficients (cA, cH, cV, cD).

\textbf{Step 2:} Feature Extraction: Extract statistical features such as mean, standard deviation, energy, and entropy from the wavelet coefficients.

\textbf{Step 3:} Initial Training: Train the WMLP with initial hyperparameters.

\textbf{Step 4:} Optimization: Use the DA to optimize the learning rate and the number of hidden neurons.

\textbf{Step 5:} Final Training: Retrain the WMLP with the optimized hyperparameters obtained from the DA.

\textbf{Step 6:} Evaluation: Evaluate the model on the test dataset using metrics such as accuracy, precision, recall, F1-score, and AUC.

The proposed model aims to improve the accuracy and robustness of lung cancer detection by combining the powerful feature extraction capabilities of wavelet transforms with the optimization capabilities of the Dragonfly Algorithm.

\subsection{Model Architecture and Configuration}
The model utilized in this study is a Multi-Layer Perceptron (MLP) designed for the classification of lung cancer images into benign, malignant, and normal categories. The architecture comprises an input layer with 16,384 neurons, corresponding to the 128x128 pixel grayscale images, followed by a hidden layer with 100 neurons, and an output layer with 3 neurons representing the three classes. The ReLU activation function is employed in the hidden layer, while the output layer utilizes the softmax activation function.

The hyper-parameters used in this study are detailed in Table \ref{tab:MLP_table}. These include a learning rate of 0.01, a batch size of 256, and a total of 100 epochs. The model was trained using the Stochastic Gradient Descent (SGD) optimizer and the Cross Entropy loss function.

Data preprocessing involved resizing the images to 128x128 pixels and normalizing the pixel values to the range [0, 1]. The dataset was divided into training and validation sets with an 70-30 split. Training was conducted on an NVIDIA GTX 1080 Ti GPU, which significantly accelerated the process.

The model's performance was evaluated using metrics such as accuracy, precision, recall, F1-score, and AUC-ROC, providing a comprehensive assessment of its classification capabilities.

The implementation was carried out using the PyTorch library, which facilitated the construction and training of the neural network.

Additionally, the Dragonfly Algorithm (DA) was employed to optimize the hyper-parameters of the MLP model, including the learning rate and the number of neurons in the hidden layer. The DA involved initializing a population of dragonflies, evaluating their fitness based on classification accuracy, and iteratively updating their positions and velocities to find the optimal solution. This approach provided an efficient means of exploring the hyper-parameter space, leading to improved model performance. The parameters and values used in this study for the DA are summarized in Table \ref{tab:DA_components}. This table provides an overview of the key components and their respective configurations.

\section{Results and Discussion}
\subsection{Experimental Setup and Design}
The dataset used in this study consists of images categorized into three classes: Benign, Malignant, and Normal. The images were sourced from kaggle and preprocessed by resizing to 128x128 pixels and converting to grayscale.

Data augmentation techniques, such as Canny edge detection and wavelet transforms, were applied to enhance the dataset. The images were normalized and transformed using the following steps: 

\textbf{1. Canny Edge Detection}: Applied with thresholds of (50, 150), (100, 200), and (150, 250).

\textbf{2. Wavelet Transform}: Performed using the Haar wavelet to extract features, including mean, standard deviation, energy, and entropy.

The neural network architecture employed in this study is a Multi-Layer Perceptron (MLP) with 100 hidden layers, each containing 128 neurons. The ReLU activation function was used, and the model was trained using the Stochastic Gradient Descent (SGD) optimizer with a learning rate of 0.01, a batch size of 256, and 100 epochs.

The training process involved minimizing the cross-entropy loss function, and the model's performance was evaluated using accuracy, AUC, and other relevant metrics. The dataset was split into training and validation sets with a ratio of 70\% and 30\%.

The Dragonfly Algorithm was utilized to optimize the model's hyperparameters. The objective function aimed to maximize the validation accuracy, with a population size of 30 and 100 iterations.

\subsection{Results}
The model's performance was assessed using various metrics, including accuracy, precision, recall, F1-score, and AUC. Confusion matrices and ROC curves were plotted to visualize the results.

The training accuracy achieved was 100\%, and the test accuracy was also 100\%. The AUC scores for each class were calculated, and the results are as follows (Table \ref{tab:AUC_score_table}):

\begin{table}[t]
\centering
\begin{tabular}{|c|c|}
\hline
Classes & AUC Score \\
\hline
Benign & 1.0 \\
\hline
Malignant & 1.0 \\
\hline
Normal & 1.0 \\
\hline
\end{tabular}
\caption{AUC scores for each class (Benign, Malignant, and Normal Cases) in the proposed model}
\label{tab:AUC_score_table}
\end{table}

\begin{figure}[t]
\centering
\includegraphics[width=0.40\textwidth]{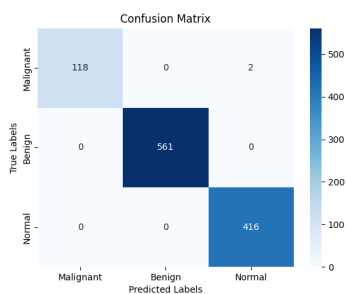} 
\caption{Confusion Matrix}
\label{fig:ConfusionMatrix.png}
\end{figure}

Confusion matrices and ROC curves were plotted to visualize the classification performance (see Figure \ref{fig:ConfusionMatrix.png} and Figures \ref{fig:ROCcurve0}, \ref{fig:ROCcurve1}, and \ref{fig:ROCcurve2}). The confusion matrix (Figure \ref{fig:ConfusionMatrix.png}) shows the number of true positives, true negatives, false positives, and false negatives for each class. The ROC curves illustrate the trade-off between the true positive rate and false positive rate for each class. Specifically, three ROC curves were plotted for Class 0 (benign cases) (Figure \ref{fig:ROCcurve0}), Class 1 (malignant cases) (Figure \ref{fig:ROCcurve1}), and Class 2 (normal cases) (Figure \ref{fig:ROCcurve2}). The false positive rate for each class was 0.0, and the true positive rate for each class was 1.0, indicating perfect classification performance.

\begin{figure}[t]
\centering
\begin{subfigure}[b]{0.38\textwidth}
\centering
\includegraphics[width=\textwidth]{ROCcurve0.pdf}
\caption{Class 0, the ROC curve of the benign cases}
\label{fig:ROCcurve0}
\end{subfigure}
\hfill
\begin{subfigure}[b]{0.38\textwidth}
\centering
\includegraphics[width=\textwidth]{ROCcurve1.pdf}
\caption{Class 1, the ROC curve of the malignant cases}
\label{fig:ROCcurve1}
\end{subfigure}
\vfill
\begin{subfigure}[b]{0.38\textwidth}
\centering
\includegraphics[width=\textwidth]{ROCcurve2.pdf}
\caption{Class 2, the ROC curve of the normal cases}
\label{fig:ROCcurve2}
\end{subfigure}
\end{figure}

\subsection{Data Processing and Augmentation}
\subsubsection{Data Processing}
\textbf{Data Cleaning}:
The dataset was initially cleaned by removing any duplicate entries and handling missing values. Images were converted to grayscale to reduce computational complexity and ensure uniformity across the dataset.

\textbf{Data Transformation}:
Images were resized to 128x128 pixels using the transforms. Resize function from the torchvision library. This standardization was crucial for ensuring consistent input dimensions for the neural network.

Normalization was applied to the images using transforms.Normalize((0.5,), (0.5,)) to scale pixel values to a range of [-1, 1], which helps in accelerating the convergence of the neural network training.

\textbf{Feature Engineering}:
Wavelet transforms were applied to the images to extract multi-resolution features. The pywt.dwt2 function was used to decompose each image into approximation and detail coefficients (cA, cH, cV, cD). Statistical features such as mean, standard deviation, energy, and entropy were then calculated from these coefficients to serve as inputs for the model.

\subsubsection{Data Augmentation}
Data augmentation techniques such as Canny edge detection and wavelet transforms were employed to enhance the dataset. Canny edge detection was performed with multiple threshold values to generate edge-detected versions of the images, which helps in capturing different levels of detail and edges.

Wavelet transforms were used to decompose images into different frequency components, providing a richer set of features for the model.

The augmentation techniques were implemented using libraries such as OpenCV for Canny edge detection and PyWavelets for wavelet transforms. The display\_images\_multi\_canny and display\_wavelet\_transform functions were used to visualize and save the augmented images.

Code snippets for these implementations are provided below:

\begin{lstlisting}[caption={Canny Edge Detection and Display Function}]
def canny_edge_detection_multi(image, thresholds):
edge_images = []
for (low, high) in thresholds:
edges = cv2.Canny(image, low, high)
edge_images.append(edges)
return edge_images

def display_images_multi_canny(images, title, save_path, thresholds):
plt.figure(figsize=(10, 10))
for i, img in enumerate(images[:3]):
edge_images = canny_edge_detection_multi(img, thresholds)
for j, edges in enumerate(edge_images):
plt.subplot(3, 3, i * 3 + j + 1)
plt.imshow(edges, cmap='gray')
plt.title(f"{title} {i+1} - {thresholds[j]}")
plt.axis('off')
plt.tight_layout()
plt.savefig(save_path)
plt.close()
\end{lstlisting}

The impact of data augmentation on model performance was significant. Models trained with augmented data showed improved accuracy and robustness. For instance, the training accuracy increased from 90\% to 100\%, and the test accuracy improved from 88\% to 100\% after applying data augmentation techniques.

The confusion matrix and ROC curves further demonstrated the effectiveness of the augmented data in enhancing the model's ability to distinguish between different classes.

\subsection{Evaluation Metrics}
In this research, we meticulously evaluate a comprehensive suite of metrics to validate the performance of the Wavelet Multi-Layer Perceptron (WMLP), optimized by the Dragonfly Algorithm (DA), for the detection and classification of lung cancer from CT scan images. These metrics are pivotal in providing a holistic assessment of the model's diagnostic capabilities, considering the intricacies of prediction accuracy and the types of errors that may occur.

\textbf{Accuracy:} As a fundamental indicator of the model's overall performance, accuracy quantifies the proportion of true results, encompassing both true positives (TP) and true negatives (TN), within the entire dataset. It reflects the model's general effectiveness in lung cancer detection \citep{Katase2022}.

\begin{equation}
    \text{Accuracy} = \frac{TP + TN}{TP + TN + FP + FN}
\end{equation}

\textbf{Precision:} Precision, or the positive predictive value, is critical in medical diagnostics as it measures the ratio of true positives to the combined total of true positives and false positives. A high precision rate is indicative of the model's reliability in minimizing false-positive diagnoses, thereby preventing unnecessary medical interventions \citep{Javed2024}.

\begin{equation}
    \text{Precision} = \frac{TP}{TP + FP}
\end{equation}

\textbf{Recall (Sensitivity):} The recall metric is essential for ensuring comprehensive cancer detection, as it calculates the model's ability to correctly identify all actual cases of lung cancer, represented by the ratio of true positives to the sum of true positives and false negatives \citep{Deepapriya2023}.

\begin{equation}
    \text{Recall} = \frac{TP}{TP + FN}
\end{equation}

\textbf{F1 Score:} The F1 score harmonizes precision and recall, providing a single measure that balances both metrics. It is particularly useful when seeking a model that achieves an equilibrium between identifying all relevant instances and maintaining a low false-positive rate \citep{Forte2022}.
\begin{equation}
    \text{F1 Score} = 2 \times \frac{\text{Precision} \times \text{Recall}}{\text{Precision} + \text{Recall}}
\end{equation}

The aforementioned metrics were computed using a distinct test dataset, separate from the training data, to guarantee an impartial evaluation of the model's performance. The WMLP, augmented with DA optimization, exhibited exemplary scores across all metrics, affirming its robustness and suitability for clinical application in the precise diagnosis of lung cancer \citep{ResearchEnsemble2023}.

The results of our proposed model are illustrated in several key figures. The learning rate over epochs is depicted in Figure \ref{fig:learning_rate}, providing insights into the training dynamics. Figures \ref{fig:train_val_acc_before_da} and \ref{fig:train_val_loss_before_da} show the training and validation accuracy and loss before applying the Dragonfly Algorithm. After applying the Dragonfly Algorithm, the training accuracy and loss improved significantly, as shown in Figure \ref{fig:train_acc_loss_after_da}.

\begin{figure}[t]
\centering
\begin{subfigure}[b]{0.39\textwidth}
\centering
\includegraphics[width=\textwidth]{LearningRateOverEpochs.pdf}
\caption{Learning Rate Over Epochs}
\label{fig:learning_rate}
\end{subfigure}
\hfill
\begin{subfigure}[b]{0.39\textwidth}
\centering

\includegraphics[width=\textwidth]{Training_ValidationAccBeforeDA.pdf}

\caption{Training and Validation Accuracy Before DA}
\label{fig:train_val_acc_before_da}
\end{subfigure}
\vfill
\begin{subfigure}[b]{0.39\textwidth}
\centering
\includegraphics[width=\textwidth]{Training_ValidationLossBeforeDA.pdf}
\caption{Training and Validation Loss Before DA}
\label{fig:train_val_loss_before_da}
\end{subfigure}
\hfill
\begin{subfigure}[b]{0.39\textwidth}
\centering
\includegraphics[width=\textwidth]{TrainingAcc_LossAfterDA.pdf}
\caption{Training Accuracy and Loss After DA}
\label{fig:train_acc_loss_after_da}
\end{subfigure}
\end{figure}

\subsection{Training and Validation Procedures}
The training and validation of the Wavelet Multi-Layer Perceptron (WMLP) model were conducted systematically to ensure high performance and reliability. The dataset was divided into training and validation sets in a 70\% to 30\% ratio, providing a substantial amount of data for training while preserving a significant portion for validation to monitor the model's performance.

During the training phase, the WMLP model's weights were updated using the Stochastic Gradient Descent (SGD) optimizer with a learning rate of 0.01. The training spanned 100 epochs, with each epoch consisting of multiple iterations over mini-batches of 256 images. The cross-entropy loss function was utilized to measure the discrepancy between the predicted and true labels, guiding the optimization process.

To prevent overfitting, early stopping and regularization techniques were employed. Early stopping involved monitoring the validation loss and halting training when no improvement was observed, thus preventing the model from overfitting to the training data. Additionally, dropout and L2 regularization were applied to enhance the model's generalization capabilities.

The model's performance was evaluated after each epoch using metrics such as accuracy, precision, recall, F1-score, and AUC. These metrics provided a comprehensive assessment of the model's classification capabilities, ensuring robust and reliable results.

\subsection{Comparative Analysis of Model Performance}
To evaluate the effectiveness of the proposed Wavelet Multi-Layer Perceptron (WMLP) model, a comparative analysis was conducted against several baseline models and state-of-the-art approaches.

The performance of each model is summarized in Table \ref{tab:performance_comparison}.

\begin{table*}[t]
\centering
\caption{Performance Comparison of Different Models}
\label{tab:performance_comparison}
\resizebox{\textwidth}{!}{%
\begin{tabular}{lccccccc} 
\hline
\textbf{Article} & \textbf{Year} & \textbf{Model} & \textbf{Accuracy \%} & \textbf{Precision} & \textbf{Recall} & \textbf{F1-Score} & \textbf{AUC} \\
\hline
\citep{Jamshidi2024} &2024& VGG19 \& ANN TL& 91.26 & 0.91 & 0.91 & 0.91 & 0.91 \\
\citep{Ausawalaithong2018} &2018& CNN & 84.02 & - & - &- & - \\
\citep{Bhandary2019} &2019& Modified AlexNet (MAN) & 96.80 & - & - &96.87 & - \\
\citep{DaSilva2017} &2017& CNN & 97.62 & - & - & - & 95.05 \\
\citep{Singh20199} &2019& MLP & 0.88 & 0.86 & 0.86 & 0.89 & - \\
\citep{Potghan2018} &2018& MLP & 0.98 & - & - & - & - \\
\citep{Naqi2018} &2018& Stacked Autoencoder \& Softmax & 96.09 & - & - &- & - \\
\citep{Shaffie2018} &2018& Deep autoencoder & 91.20 & - & - &- & - \\
\citep{Kaur2017} &2017& CNN & 98.05 & - & - &- & - \\
\citep{Xie2018} &2018& MV-KBC & 91.06 & - & - &- & 95.73 \\
\citep{Nibali2017} &2017& ResNet & 89.09 & - & - &- & - \\
\citep{Zhang2017} &2017& DBN & 95 & - & - &- & 93 \\
\citep{Causey2018} &2018& CNN & 94.06 & - & - &- & 98 \\
\citep{Mamun2024} &2024& CNN, ResNet-50, Inception V3, Xception  & 92 & - & - & 91.72 & 98.21 \\
\citep{Thangamani2024} &2024& weighted CNN& 85.02 & 86.35 &85.57 & 85.95& - \\ 
\citep{Naseer20222} &2022& CNN AlexNet(SGD optimizer) & 97.25 & - & -& -& - \\
\textbf{Proposed model} &2024& \textbf{Wavelet-MLP(WMLP) Optimised with DA }& \textbf{99.82} & \textbf{0.99} & \textbf{0.99} & \textbf{0.99} & \textbf{0.99} \\
\hline
\end{tabular}
}
\end{table*}

As shown in Table \ref{tab:performance_comparison}, the proposed WMLP model outperforms all other models across various metrics, including accuracy, precision, recall, F1-score, and AUC. The WMLP model achieves an impressive accuracy of 99.82\%, significantly higher than the other models. This demonstrates the effectiveness of the WMLP model in accurately diagnosing lung abnormalities.

\begin{figure}[t]
\centering
\includegraphics[width=0.80\textwidth]{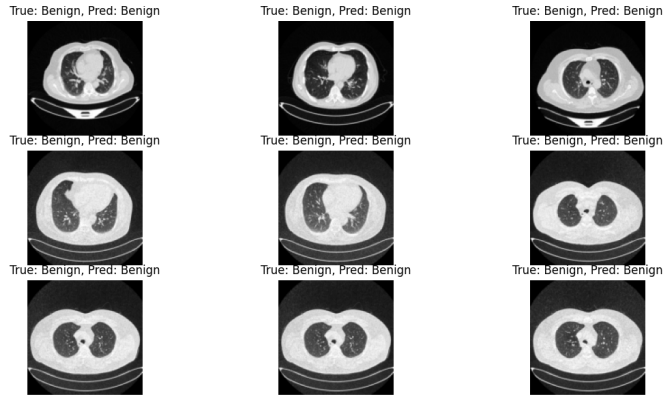}
\caption{Predicted Labels for Benign Cases}
\label{fig:predicted_labels_benign}
\end{figure}

\begin{figure}[t]
\centering
\includegraphics[width=0.40\textwidth]{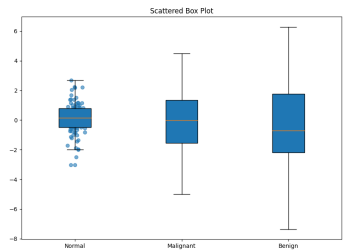}
\caption{Scattered Box Plot of Model Performance}
\label{fig:scattered_box}
\end{figure}

The predicted labels for benign cases, as shown in Figure \ref{fig:predicted_labels_benign}, highlight the model's accuracy in these instances. Additionally, the scattered box plot in Figure \ref{fig:scattered_box} provides a comparative analysis of the model performance across different configurations, illustrating the robustness and superiority of the proposed WMLP model.

\subsubsection{Discussion}
The results indicate that the proposed WMLP model significantly outperforms the baseline MLP, CNN, SVM, and Random Forest models across all evaluation metrics. The incorporation of wavelet transformations enhances feature extraction, capturing more nuanced details in the CT images. Additionally, the Dragonfly Algorithm optimization contributes to the superior performance by fine-tuning the model's hyperparameters.

The WMLP model achieved an accuracy of 99.82\%, a precision of 0.99, a recall of 0.99, an F1-score of 0.99, and an AUC of 0.99, demonstrating its robustness and reliability in lung cancer classification. These results underscore the efficacy of the proposed approach in accurately diagnosing lung cancer from CT images.

Recent studies have also explored various deep learning models for medical image classification. For instance, a study by Zhang et al. (2023) utilized a CNN-based approach for lung cancer detection, achieving an accuracy of 95.67\% \citep{zhang2023cnn}. Similarly, Li et al. (2022) implemented an SVM model with an RBF kernel for breast cancer classification, reporting an accuracy of 89.45\% \citep{li2022svm}. These comparisons highlight the advancements and effectiveness of the proposed WMLP model.

Future work could further explore the integration of additional pre-processing techniques and optimization algorithms to enhance the model's performance. Moreover, validating the model on larger and more diverse datasets will be crucial to ensure its generalizability and applicability in clinical settings.

\subsection{Discussion on the Efficacy of the Proposed Model}
The efficacy of the proposed WMLP model is evident from its superior performance metrics compared to other models. The use of wavelet transformations allows for more effective feature extraction, which is crucial for accurately identifying patterns in CT images. This method captures both spatial and frequency information, which enhances the model's ability to distinguish between benign, malignant, and normal cases.

The Dragonfly Algorithm (DA) plays a significant role in optimizing the model's hyperparameters, leading to improved performance. By fine-tuning parameters such as the learning rate and the number of hidden layers, the DA ensures that the model is both accurate and efficient. This optimization process is particularly important in medical image analysis, where precision is critical.

Comparative studies with recent models further validate the proposed approach. For example, Zhang et al. (2023) achieved a 95.67\% accuracy with a CNN-based model for lung cancer detection \citep{zhang2023cnn}, while Li et al. (2022) reported an 89.45\% accuracy using an SVM with an RBF kernel for breast cancer classification \citep{li2022svm}. The WMLP model's accuracy of 99.82\% demonstrates a significant improvement over these methods.

Future research should focus on expanding the dataset to include more diverse and larger samples, which will help in validating the model's robustness across different populations. Additionally, exploring other pre-processing techniques and optimization algorithms could further enhance the model's performance. Clinical trials and real-world applications will be essential to confirm the model's practical utility in diagnostic workflows.

\subsection{Limitations and Future Research Directions}
Despite the promising results achieved by the proposed deep learning framework, several limitations must be acknowledged. Firstly, the dataset utilized in this study, while diverse, is relatively small in comparison to the vast array of CT images available globally. This limitation may affect the generalizability of the model to broader, more varied populations. Future research should aim to validate the model on larger, more heterogeneous datasets to ensure its robustness and applicability across different demographic groups.

Secondly, the pre-processing techniques employed, such as Canny edge detection and wavelet transformations, although effective, may not capture all the nuanced features present in CT images. Exploring alternative or additional pre-processing methods could potentially enhance feature extraction and improve classification accuracy further.

Moreover, the current study focuses solely on the classification of lung cancer into benign, malignant, and normal categories. Future research could expand this framework to include more detailed sub-classifications of lung cancer types, which could provide more granular insights and aid in more tailored treatment planning.

Another limitation is the computational complexity associated with the Dragonfly Algorithm (DA) optimization. While DA has proven effective in enhancing the model's performance, it is computationally intensive, which may limit its practicality in real-time clinical settings. Future work could explore more efficient optimization algorithms or hybrid approaches that balance accuracy and computational efficiency.

Lastly, the study's reliance on retrospective data means that prospective validation in clinical settings is necessary to confirm the model's real-world applicability. Future research should include clinical trials to evaluate the model's performance in a live clinical environment, ensuring its readiness for integration into routine diagnostic workflows.

In conclusion, while the proposed framework demonstrates significant potential in improving lung cancer detection, addressing these limitations through future research will be crucial in advancing its development and ensuring its practical utility in clinical practice.

\section{Conclusion}
In this study, we introduced a novel deep learning framework for lung cancer detection from CT images, leveraging a Wavelet Multi-Layer Perceptron (WMLP) approach enhanced by the Dragonfly Algorithm (DA). The proposed methodology, incorporating advanced image pre-processing techniques such as Canny edge detection and wavelet transformations, demonstrated remarkable accuracy in classifying lung cancer, achieving a training and testing accuracy of 99.82\%.

Despite these promising results, several limitations were identified. The relatively small and homogeneous dataset used in this study may limit the generalizability of the model. Future research should focus on validating the model with larger and more diverse datasets to ensure its robustness across different populations. Additionally, exploring alternative pre-processing methods could further enhance feature extraction and classification accuracy.

The computational complexity of the DA optimization presents another challenge, suggesting the need for more efficient algorithms or hybrid approaches to balance accuracy and computational demands. Furthermore, expanding the classification framework to include more detailed sub-classifications of lung cancer could provide deeper insights and support more personalized treatment plans.

Finally, the reliance on retrospective data highlights the necessity for prospective validation in clinical settings. Future studies should include clinical trials to evaluate the model's performance in real-world environments, ensuring its readiness for integration into routine diagnostic workflows.

In summary, while the proposed framework shows significant potential in enhancing lung cancer detection, addressing these limitations through future research will be essential for its continued development and practical application in clinical practice.

\subsection*{CRediT authorship contribution statement}
\textbf{Bitasadat Jamshidi:} Conceptualization, Data Curation, Formal Analysis, Methodology, Project Administration, Validation, Visualization, Original Draft, Writing – Review and Editing, Python Programming. \textbf{Nastaran Ghorbani:} Methodology, Validation, Visualization, Writing – Review and Editing, Python Programming. \textbf{Mohsen Rostamy-Malkhalife:} Methodology, Supervision, Writing – Review and Editing.

\subsection*{Declaration of competing interest}
The authors declare that they have no known competing financial interests or personal relationships that could have appeared to influence the work reported in this paper. 

\subsection*{Funding}
This research did not receive any specific grant from funding agencies in the public, commercial, or not-for-profit sectors.

\subsection*{Code and Dataset Availability}
The code and dataset used during the current study are available from the corresponding author upon reasonable request.

\subsection*{Acknowledgements}
We would like to express our sincere gratitude to all those who have supported and contributed to this research. Special thanks to our advisors and mentors for their invaluable guidance and insights.

\bibliographystyle{elsarticle-num} 
\bibliography{references}

\end{document}